\documentclass{amsart}
\usepackage{amssymb}

\usepackage{graphicx}
\usepackage{amscd}

\newtheorem{theorem}{Theorem}
\theoremstyle{plain}

\newtheorem{corollary}{Corollary}

\newtheorem{lemma}{Lemma}

\newtheorem{proposition}{Proposition}

\numberwithin{equation}{section}
\numberwithin{theorem}{section}
\numberwithin{lemma}{section}
\numberwithin{proposition}{section}
\numberwithin{corollary}{section}

\input{tcilatex}

\begin{document}
\title[$p-$adic pseudo-differential equations]{pseudo-differential equations connected \ with $p-$adic forms and local zeta
functions}
\author{W. A. Zuniga-Galindo}
\address{Department of Mathematics and Computer Science, Barry University, 11300 N.E.
Second Avenue, Miami Shores, Florida 33161, USA}

\begin{abstract}
We study the asymptotics of fundamental solutions of $p$-adic
pseudo-differential equations of type 
\begin{equation*}
\left( f(\partial ,\beta )+\lambda \right) u=g,\ 
\end{equation*}
where $f(\partial ,\beta )$ is a pseudo-differential operator with symbol $%
\left| f\right| _{K}^{\beta }$, $\beta >0$, $f$ is a form of arbitrary
degree with coefficients in a $p-$adic field, $\lambda \geq 0$, $\ $and $g$
\ is a Schwartz-Bruhat function.
\end{abstract}

\email{wzuniga@mail.barry.edu}
\maketitle

\section{Introduction}

Let $K$ be a $p-$adic field, i.e. $\left[ K:\mathbb{Q}_{p}\right] <\infty $.
Let $R_{K}$ be the valuation ring of $K$, $P_{K}$ the maximal ideal \ of \ $%
R_{K}$, and \ $\overline{K}=R_{K}/$ $P_{K}$ the residue field \ of $K$. The
cardinality of $\overline{K}$ is denoted by $q$. For $z\in K$, $v(z)\in 
\mathbb{Z}\cup \left\{ +\infty \right\} $ denotes the valuation of $z$, $%
\left| z\right| _{K}=q^{-v(z)}$ and $ac$ $z=z\frak{p}^{-v(z)}$ where \ $%
\frak{p}$ is a fixed uniformizing parameter for \ $R_{K}$. For $x=\left(
x_{1},\ldots ,x_{n}\right) \in K^{n}$ we put $\left\| x\right\|
_{K}=\max_{1\leq i\leq n}\left| x_{i}\right| _{K}$.

We denote by $S(K^{n})$ the $\mathbb{C}$-vector space of Schwartz-Bruhat
functions over $K^{n}$. The dual \ space $S^{\prime }(K^{n})$ is the space
of distributions over $K^{n}$. Let \ $f=f\left( x\right) \in K\left[ x\right]
$, $x=\left( x_{1},\ldots ,x_{n}\right) $, be a non-constant polynomial, and 
$\beta $ a positive \ real number. A $p$-adic pseudo-differential operator $%
f(\partial ,\beta )$, with symbol $\left| f\right| _{K}^{\beta }$, is an
operator of the form 
\begin{equation*}
\begin{array}{cccc}
f(\partial ,\beta ): & \mathcal{S}(K^{n}) & \rightarrow & \mathcal{S}(K^{n})
\\ 
& \Phi & \rightarrow & \mathcal{F}^{-1}\left( \left| f\right| _{K}^{\beta }%
\mathcal{F}\left( \Phi \right) \right) ,
\end{array}
\end{equation*}
where $\mathcal{F}$\ \ is the Fourier transform. The operator \ $f(\partial
,\beta )$ is continuous \ and \ has self-adjoint extension with dense domain
in $L^{2}\left( K^{n}\right) $. This operator is considered to be a $p-$adic
analogue of a linear partial differential operator with constant
coefficients. The $p$-adic pseudo-differential equation \ 
\begin{equation}
f(\partial ,\beta )u=g\text{, \ }g\in \mathcal{S}(K^{n})\text{,}  \label{eq1}
\end{equation}
is naturally associate to $f(\partial ,\beta )$. The theory of $p-$adic
pseudo-differential equations is emerging \ \ motivated \ by the use of $p-$%
adic models in physics \cite{Koch1}, \cite{VVZ}. The state of the art of the
theory of $p-$adic pseudo-differential operators is exposed in a recent book
by Kochubei \cite{Koch1}. The simplest possible \ operator has symbol $%
\left| x\right| _{K}^{\beta }$, $\beta >0$. Vladimirov studied extensively
this class of operators showing, among other results, the existence of
fundamental solutions \cite{V}, \cite{VVZ}. Kochubei showed the existence of
fundamental solutions for elliptic operators, i.e., for operators with
symbols \ of the form $\left| f\left( x_{1},\ldots ,x_{n}\right) \right|
_{K}^{\beta }$, $\beta >0$, \ where \ $f\left( x_{1},\ldots ,x_{n}\right) $
is a quadratic form satisfying $f\left( x_{1},\ldots ,x_{n}\right) \neq 0$ \
when $\left| x_{1}\right| _{K}+..+\left| x_{n}\right| _{K}\neq 0$, \cite
{Koch1}, \cite{Koch2}. In \cite{Khr} Khrennikov considered spaces of
functions and distributions defined outside \ the singularities \ of a
symbol, in this situation he showed the existence of a fundamental solution
for a $p-$adic pseudo-differential equation with symbol $\ a\left( \xi
\right) \neq 0$. In a recent note \ \cite{Z} the author \ observed that
Atiyah's proof \cite{A} for the existence of a fundamental solution for a
linear partial differential operator with constant coefficients can be
adapted to the $p-$adic case. In this proof \ the meromorphic continuation
of the Igusa local zeta function plays a central role. On the other hand,
Jang \cite{J} and Sato \cite{S} showed \ explicitly a connection between the
local zeta function of a quadratic form and the $p-$adic Green function\ $%
G_{\lambda }$ \ (a fundamental solution) of the following
pseudo-differential equation 
\begin{equation}
\left( f\left( \partial ,\beta \right) +\lambda \right) u=g,\ \lambda >0,\
g\in \mathcal{S}(K^{n}),  \label{eq2a}
\end{equation}
when $\beta =1$, and $f$\ is a relative invariant of some prehomogeneous
vector space. In \cite{S} Sato \ showed that the asymptotics of $G_{\lambda
} $, as $\left| f\right| _{K}\rightarrow \infty $, is controlled by the
gamma factor of the functional equation \ of the local zeta function
associated to $f$. Previously, Kochubei \cite{K2}, \cite{K3},\ had described
the asymptotics of $G_{\lambda }$, at the infinity and at the origin, \ when 
$f\left( \partial ,\beta \right) $\ is an elliptic operator.

This paper is dedicated to the study of the asymptotics of fundamental
solutions for (\ref{eq1}) and (\ref{eq2a}) when $f$ is a homogeneous
polynomial in an arbitrary number of variables. We construct fundamental
solutions ``at infinity and at the origin'' for (\ref{eq1}), that is,
solutions of (\ref{eq1} ) \ when $g$ is the characteristic function of a
ball around the infinity, respectively, around the origin (cf. Theorem \ref
{th1}, and Corollary \ref{corola1}). The proof of Theorem \ref{th1} uses
resolution of singularities \cite{H}, \ and \ some ideas developed by Atiyah
for solving \ the problem of division of a distribution by an analytic
function \cite{A}. The techniques used in the proof of Theorem \ref{th1}
allow us to establish a functional equation for \ the distributions of type $%
\chi (ac$ $f)\left| f\right| _{K}^{s}$ (cf. Theorem \ref{thfe}) on a certain
subspace of $\mathcal{S}(K^{n})$. Functional equations for distributions of
type $\chi \left( ac\text{ }f\right) \left| f\right| _{K}^{s}$ have been
obtained by Igusa \cite{I4}, \ Sato \cite{S1}, Gyoja \cite{G}, and Denef and
Meuser \cite{DM}.

We also give the asymptotic expansion of the Green function $G_{\lambda }(x)$
as $\left\| x\right\| _{K}\rightarrow \infty $ (cf. Theorem \ref{th4}). The
proof of this result uses the technique of integration on the fibers and
some properties of the Igusa local zeta function. Kochubei \ studied the
asymptotics of the Green functions associated with elliptic operators 
\cite[Sect. 2.8]{Koch1}, \cite{K2}, \cite{K3}, at infinity and at the
origin. The asymptotics obtained by Kochubei at infinity can be recovered
from \ Theorem \ref{th4}.\ 

\section{Preliminaries}

Let $\Psi :K\rightarrow \mathbb{C}^{\times }$ be the additive character
defined by 
\begin{equation*}
\begin{array}{cccccccc}
\Psi : & K & \rightarrow & \mathbb{Q}_{p} & \rightarrow & \mathbb{Q}_{p}/%
\mathbb{Z}_{p} & \rightarrow & \mathbb{C}^{\times } \\ 
& x & \rightarrow & Tr_{K/\mathbb{Q}_{p}}\left( x\right) &  & y & \rightarrow
& \exp \left( 2\pi iy\right) ,
\end{array}
\end{equation*}
where $Tr_{K/\mathbb{Q}_{p}}$\ is the trace \ mapping. Let $\left| dx\right| 
$\ \ denote the Haar measure on $K^{n}$ \ normalized such that $%
vol(R_{K}^{n})=1$. We shall identify the $n-$dimensional $K-$vector space $%
K^{n}$ with its dual vector space via the standard inner product 
\begin{equation*}
\left[ x,y\right] =\sum_{i=1}^{n}x_{i}y_{i}\text{, \ }x\text{, }y\in K^{n}.
\end{equation*}
The Haar measure $\left| dx\right| $ is autodual with respect the pairing $%
\Psi \left( \left[ x,y\right] \right) $. \ For $\Phi \in S(K^{n})$, the
Fourier transform $\mathcal{F}\Phi $ of $\Phi $\ is defined by 
\begin{equation*}
\left( \mathcal{F}\Phi \right) \left( \xi \right) =\int\limits_{K^{n}}\Phi
\left( x\right) \Psi \left( -\left[ x,\xi \right] \right) \left| dx\right| .
\end{equation*}
The Fourier transform induces a linear isomorphism of $S(K^{n})$ onto
itself, and the inverse transform is given by 
\begin{equation*}
\Phi \left( x\right) =\int\limits_{K^{n}}\left( \mathcal{F}\Phi \right)
\left( \xi \right) \Psi \left( \left[ x,\xi \right] \right) \left| d\xi
\right| .
\end{equation*}
The Fourier transform can be extended to an isometry of $L^{2}\left(
K^{n}\right) $ onto $L^{2}\left( K^{n}\right) $.

We denote by $S^{\prime }(K^{n})$ the dual space of $S(K^{n})$,\ i.e. the
space of complex valued distributions on $K^{n}$. If $T\in S^{\prime
}(K^{n}) $, we denote by $\mathcal{F}T$\ its Fourier transform, that is the
distribution $\left\langle \mathcal{F}T\ ,\Phi \right\rangle $ $%
=\left\langle T,\mathcal{F}\Phi \right\rangle $.

\subsection{The Riesz kernel}

We shall collect some well-known results about the Riesz kernel that will be
used in the next sections \cite{T}, \cite{VVZ}.

The $p-$adic Gamma function $\Gamma _{n}\left( \alpha \right) $ is defined
as follows 
\begin{equation*}
\Gamma _{n}\left( \alpha \right) =\frac{1-q^{\alpha -n}}{1-q^{-\alpha }}%
\text{, }\alpha \in \mathbb{C},\alpha \neq 0\text{.}
\end{equation*}
The \ Gamma function is meromorphic with simple zeros at $n+\frac{2\pi i}{%
\log q}Z$\ and \ unique simple pole at $\alpha =0$. In addition, it
satisfies 
\begin{equation*}
\Gamma _{n}\left( \alpha \right) \Gamma _{n}\left( n-\alpha \right) =1\text{%
, }\alpha \notin \left\{ 0\right\} \cup \left\{ n+\frac{2\pi i}{\log q}z%
\text{ };\text{ }z\in \mathbb{Z}\right\} .
\end{equation*}

The Riesz kernel $R_{\alpha }$ is the distribution \ determined by the
function 
\begin{equation*}
\mathcal{R}_{\alpha }\left( x\right) =\frac{\left\| x\right\| _{K}^{\alpha
-n}}{\Gamma _{n}\left( \alpha \right) }\text{, \ }x\in K^{n}\text{, }\func{Re%
}(\alpha )>0\text{, }\alpha \notin n+\frac{2\pi i}{\log q}\mathbb{Z}.
\end{equation*}

The Riesz kernel possesses, as a distribution, a meromorphic continuation \
to $\mathbb{C}$ given by 
\begin{eqnarray}
\left\langle \mathcal{R}_{\alpha },\Phi \right\rangle &=&\Phi \left(
0\right) \left( \frac{1-q^{-n}}{1-q^{\alpha -n}}\right) +\left( \frac{%
1-q^{-\alpha }}{1-q^{\alpha -n}}\right) \int\limits_{\left\| x\right\|
_{K}\leq 1}\left( \Phi \left( x\right) -\Phi \left( 0\right) \right) \left\|
x\right\| _{K}^{\alpha -n}\left| dx\right|  \notag \\
&&+\left( \frac{1-q^{-\alpha }}{1-q^{\alpha -n}}\right) \int\limits_{\left\|
x\right\| _{K}>1}\Phi \left( x\right) \left\| x\right\| _{K}^{\alpha
-n}\left| dx\right| ,  \label{10}
\end{eqnarray}
with poles at $n+\frac{2\pi i}{\log q}\mathbb{Z}$. We note that $%
\left\langle \mathcal{R}_{\alpha },\Phi \right\rangle \mid _{\alpha
=0}=\left\langle \delta ,\Phi \right\rangle $, i.e. $\frac{1}{\Gamma
_{n}\left( 0\right) }\left\| x\right\| _{K}^{-n}$ \ is equal to the Dirac
delta function.

\begin{proposition}[{\protect\cite[Theorem 4.5]{T}}]
\label{prop1}As elements of \ $S^{\prime }(K^{n})$, 
\begin{equation*}
\mathcal{F}\left( \frac{\left\| x\right\| _{K}^{\alpha -n}}{\Gamma
_{n}\left( \alpha \right) }\right) =\left\| x\right\| _{K}^{-\alpha },\text{
\ }\alpha \notin n+\frac{2\pi i}{\log q}\mathbb{Z}.
\end{equation*}
\end{proposition}

\subsection{Igusa's local zeta function}

Let $g(x)\in K\left[ x\right] $, $x=\left( x_{1},\ldots ,x_{n}\right) $, be
a non-constant polynomial, the $p-$adic complex power $\left| g\right|
_{K}^{s}$ associated to $g$\ (also called the Igusa local zeta function of $%
g $) is the distribution 
\begin{equation}
\left\langle \left| g\right| _{K}^{s},\Phi \right\rangle
=\int\limits_{K^{n}\setminus g^{-1}\left( 0\right) }\Phi \left( x\right)
\left| g\left( x\right) \right| _{K}^{s}\left| dx\right| \text{, }s\in 
\mathbb{C}\text{, }\func{Re}(s)>0\text{.}  \label{for3}
\end{equation}
For a fixed $\Phi $ we put $Z_{\Phi }\left( s,g\right) =\left\langle \left|
g\right| _{K}^{s},\Phi \right\rangle $, $\func{Re}(s)>0$. In the case in
which $\Phi $ is the characteristic function of \ $R_{K}^{n}$ we denote the
local zeta function of $g$ \ by $Z\left( s,g\right) $. The local zeta
functions were introduced by Weil \cite{W} and their basic properties for
general $g$ were first studied by Igusa \cite{I1}, \cite{I2}.

A basic tool in the study of the local zeta functions is Hironaka's
resolution Theorem \cite{H}. This theorem \ guarantees the existence \ of an 
$n-$dimensional $K-$analytic manifold $Y$, a finite set $E=\left\{ E\right\} 
$ of closed submanifolds of $Y$ of codimension $1$ with a pair of positive
integers $\left( N_{E},n_{E}\right) $ assigned to each $E$, and a proper $K-$%
analytic map \ $h:Y\rightarrow K^{n}$ satisfying the following properties:
(I) $\left( g\circ h\right) ^{-1}\left( 0\right) =\cup _{E\in \mathcal{E}}E$%
; (II) the restriction of $h$ to \ $Y$ $\setminus h^{-1}\left( g^{-1}\left(
0\right) \right) $ is an isomorphism onto its image; and (III) at every
point $b$ of $Y$ if $E_{1},\ldots ,E_{p}$ are all the $E$ in $\mathcal{E}$ \
containing $b$ with respective local equations $y_{1},\ldots ,y_{p}$ around $%
b$ and $\left( N_{i},n_{i}\right) =\left( N_{E},n_{E}\right) $ for $E=E_{i}$%
, then \ there exist local coordinates of $Y$ around $b$ of the form $\left(
y_{1},\ldots ,y_{p},y_{p+1},\ldots ,y_{n}\right) $ such that 
\begin{equation*}
g\circ h=\epsilon \prod\limits_{1\leq i\leq p}y_{i}^{N_{i}}
\end{equation*}
on some neighborhood of $b$, with $\epsilon $\ a unit of the local ring of $%
Y $ at $b$. The pair $(h,Y)$ is called a resolution of singularities for $%
g^{-1}(0)$, and $\cup _{E\in \mathcal{E}}\left\{ \left( N_{E},n_{E}\right)
\right\} $\ is the set of numerical data of $h$. \ A central result in the
theory of \ local zeta functions is the following.

\begin{theorem}[{Igusa, \protect\cite[Theorem 8.2.1]{I1}}]
\label{th2}Let $g(x)\in K\left[ x\right] $ be a non-constant polynomial. The
\ distribution $\left| g\right| _{K}^{s}$ admits a meromorphic continuation
\ to the complex plane such that \ $\left\langle \left| g\right|
_{K}^{s},\Phi \right\rangle $ is a rational function of $q^{-s}$ for each $%
\Phi \in S(K^{n})$. Furthermore, if \ $h:Y\rightarrow K^{n}$ is a resolution
of singularities of $g^{-1}(0)$, with numerical data $\cup _{E\in \mathcal{E}%
}\left\{ \left( N_{E},n_{E}\right) \right\} $, then 
\begin{equation*}
\prod\limits_{E\in \mathcal{E}}\left( 1-q^{-n_{E}-N_{E}s}\right) \left|
g\right| _{K}^{s}
\end{equation*}
is a holomorphic distribution. In particular the real parts of the poles \
of \ $\left| g\right| _{K}^{s}$ \ are negative rational numbers.
\end{theorem}

We shall denote \ by $\left| g\right| _{K}^{s}$ the meromorphic continuation
\ of \ distribution (\ref{for3}), and \ by $Z(s,g)$, the integral $%
\int_{R_{K}^{n}}\left| g\left( x\right) \right| _{K}^{s}\left| dx\right| $, $%
\func{Re}(s)>0$, and its\ meromorphic continuation to the complex plane.
Theorem \ref{th2} is valid for distributions of the form $\chi \left( ac%
\text{ }g\right) \left| g\right| _{K}^{s}$, where $\chi $ is a
multiplicative character of $R_{K}^{\times }$ (cf. \cite[Theorem 8.2.1]{I1}).

We set $\Omega _{l}$, $l\in \mathbb{Z},$\ for the characteristic function of
the ball $(P_{K}^{l})^{n}$. We denote by $\Delta _{0}(K^{n})$ the $\mathbb{C}
$-vector space \ generated by $\Omega _{l}$, $l\in \mathbb{N}$, \ by $\Delta
_{\infty }(K^{n})$ the $\mathbb{C}$-vector space \ generated by $\Omega
_{-l} $, $l\in \mathbb{N}$, and by $\Delta (K^{n})$ the $\mathbb{C}$-vector
space $\Delta _{0}(K^{n})\oplus \Delta _{\infty }(K^{n})$. The Fourier
transform establishes a $\mathbb{C}$-isomorphism between $\Delta _{0}(K^{n})$
and \ $\Delta _{\infty }(K^{n})$, and therefore a $\mathbb{C}$-isomorphism
from $\Delta (K^{n})$ \ onto itself.

\begin{lemma}
\label{lemma1}Let $f(x)\in K\left[ x\right] $ be a form of degree $d$. Then
the\ distribution $\left| f\right| _{K}^{s}$ \ satisfies 
\begin{equation*}
\left\langle \left| f\right| _{K}^{s},\Phi \right\rangle =\left( \frac{%
1-q^{ds}}{1-q^{-n}}\right) Z(s,f)\left\langle \mathcal{R}_{ds+n},\text{ }%
\Phi \right\rangle ,
\end{equation*}
for $s\in \mathbb{C}$, and $\Phi \in \Delta (K^{n})$.
\end{lemma}

\begin{proof}
Every $\Phi $ $\in $ $\Delta (K^{n})$ is a finite linear combination of the
form 
\begin{equation*}
\Phi (x)=\sum\limits_{i}c_{i}\Omega _{l_{i}}\left( x\right) ,
\end{equation*}
where $c_{i}\in \mathbb{C}$, $l_{i}\in \mathbb{Z}$,\ and $\Omega _{l_{i}}$
is the characteristic function of the ball $\left( P_{K}^{l_{i}}\right) ^{n}$%
. The action of $\left| f\right| _{K}^{s}$ on $\mathcal{F}\Phi $ can be
explicitly described as follows: 
\begin{equation}
\left\langle \left| f\right| _{K}^{s},\mathcal{F}\Phi \right\rangle
=\sum\limits_{i}c_{i}\left\langle \left| f\right| _{K}^{s},q^{-nl_{i}}\Omega
_{-l_{i}}\right\rangle ,  \label{11}
\end{equation}
and since 
\begin{equation}
\left\langle \left| f\right| _{K}^{s},q^{-nl_{i}}\Omega
_{-l_{i}}\right\rangle =q^{-nl_{i}}\int_{K^{n}}\left| f\left( x\right)
\right| _{K}^{s}\Omega _{-l_{i}}\left( x\right) \left| dx\right|
=Z(s,f)q^{dl_{i}s},  \label{12a}
\end{equation}
for $\func{Re}\left( s\right) >0$, it follows from (\ref{11}) that 
\begin{equation}
\left\langle \left| f\right| _{K}^{s},\mathcal{F}\Phi \right\rangle
=Z(s,f)\sum\limits_{i}c_{i}q^{dl_{i}s}\text{, \ for }\func{Re}\left(
s\right) >0.  \label{12}
\end{equation}
On the other hand, 
\begin{equation}
\left\langle \frac{1-q^{ds}}{1-q^{-n}}\mathcal{R}_{ds+n},q^{-nl_{i}}\Omega
_{-l_{i}}\right\rangle =\left\langle \frac{1-q^{-n-ds}}{1-q^{-n}}\left\|
x\right\| _{K}^{ds},q^{-nl_{i}}\Omega _{-l_{i}}\right\rangle =q^{dl_{i}s},
\label{13}
\end{equation}
for every $l_{i}\in \mathbb{Z}$, and $\func{Re}(s)>0$. Then (\ref{12}) and (%
\ref{13}) imply that

\begin{equation}
\left\langle \left| f\right| _{K}^{s},\mathcal{F}\Phi \right\rangle =\left( 
\frac{1-q^{ds}}{1-q^{-n}}\right) Z(s,f)\left\langle \mathcal{R}_{ds+n},\text{
}\mathcal{F}\Phi \right\rangle ,  \label{14}
\end{equation}
\ for $\func{Re}(s)>0$. By Theorem \ \ref{th2} and (\ref{10}), $\left|
f\right| _{K}^{s}$, $Z(s,f)$, and $\mathcal{R}_{ds+n}$ have a meromorphic
continuation to the complex plane, therefore (\ref{14}) extends to $\mathbb{C%
}$. Finally, since the Fourier transform establishes a $\mathbb{C}$%
-isomorphism from $\Delta (K^{n})$ onto itself, it is possible to remove the
Fourier transform symbol in (\ref{14}).
\end{proof}

\begin{theorem}
\label{thfe}Let $f(x)\in K\left[ x\right] $ be a non-constant form of degree 
$d$. Then the\ distribution $\left| f\right| _{K}^{s}$ \ satisfies 
\begin{equation}
\left\langle \left| f\right| _{K}^{s},\Phi \right\rangle =\frac{Z(s,f)}{Z(-s-%
\frac{n}{d},f)}\left\langle \left| f\right| _{K}^{-s-\frac{n}{d}},\mathcal{F}%
\Phi \right\rangle \text{, \ }s\in \mathbb{C\setminus }\left\{ 0\right\} 
\text{, }\Phi \in \Delta (K^{n}).  \label{11a}
\end{equation}
\end{theorem}

\begin{proof}
Suppose that $ds\notin -n+\frac{2\pi i}{\log q}\mathbb{Z}$, by rewriting (%
\ref{14}) \ as 
\begin{equation}
\left\langle \left| f\right| _{K}^{s},\mathcal{F}\Phi \right\rangle =\left( 
\frac{1-q^{-n-ds}}{1-q^{-n}}\right) Z(s,f)\left\langle \left\| x\right\|
_{K}^{ds},\text{ }\mathcal{F}\Phi \right\rangle ,\text{ }s\in \mathbb{C},
\label{16}
\end{equation}
and applying Proposition \ref{prop1} we obtain that 
\begin{equation}
\left\langle \left| f\right| _{K}^{s},\mathcal{F}\Phi \right\rangle =\left( 
\frac{1-q^{ds}}{1-q^{-n}}\right) Z(s,f)\left\langle \left\| x\right\|
_{K}^{-ds-n},\text{ }\Phi \right\rangle ,  \label{17}
\end{equation}
for $s\in \mathbb{C}\setminus \left\{ -n+\frac{2\pi i}{\log q}\mathbb{Z}%
\right\} $. By making $s\rightarrow -\left( s+\frac{n}{d}\right) $ in (\ref
{17}), \ it takes \ the following form: 
\begin{equation}
\left\langle \left| f\right| _{K}^{-s-\frac{n}{d}},\mathcal{F}\Phi
\right\rangle =\left( \frac{1-q^{-ds-n}}{1-q^{-n}}\right) Z(-s-\frac{n}{d}%
,f)\left\langle \left\| x\right\| _{K}^{ds},\text{ }\Phi \right\rangle ,
\label{18}
\end{equation}
for $s\in \mathbb{C}\setminus \left\{ \frac{2\pi i}{\log q}\mathbb{Z}%
\right\} $. By comparing (\ref{18}) and (\ref{16}), we obtain (\ref{11a})
for $s\in \mathbb{C}\setminus \left\{ \frac{2\pi i}{\log q}\mathbb{Z}%
\right\} $. In order to complete the proof, we have to show that (\ref{11a}%
)\ is valid \ for $\ s\in \left\{ \frac{2\pi i}{\log q}\mathbb{Z}\right\} $.
We recall that $R_{0}=\delta $, i.e. $\frac{1}{\Gamma _{n}\left( 0\right) }%
\left\| x\right\| _{K}^{-n}$ \ is equal to the Dirac delta function. In the
case in which $ds+n=0$ mod $\frac{2\pi i}{\log q}\mathbb{Z}$, (\ref{14})
takes the form 
\begin{eqnarray}
\left\langle \left| f\right| _{K}^{-\frac{n}{d}},\mathcal{F}\Phi
\right\rangle  &=&Z(-\frac{n}{d},f)\left\langle \mathcal{R}_{0},\text{ }%
\mathcal{F}\Phi \right\rangle =Z(-\frac{n}{d},f)\left\langle \mathcal{\delta 
},\text{ }\mathcal{F}\Phi \right\rangle   \notag \\
&=&Z(-\frac{n}{d},f)\left\langle \mathcal{F}^{-1}\delta ,\text{ }\Phi
\right\rangle =Z(-\frac{n}{d},f)\left\langle 1,\text{ }\Phi \right\rangle .
\label{19}
\end{eqnarray}

On the other hand, since $\left| f\right| _{K}^{s}$ is holomorphic at zero,
the Lebesgue Lemma implies that 
\begin{equation}
\left\langle \left| f\right| _{K}^{0},\Phi \right\rangle =\lim_{s\rightarrow
0}\left\langle \left| f\right| _{K}^{s},\Phi \right\rangle
=\int\limits_{K^{n}}\Phi \left( x\right) \left| dx\right| =\left\langle
1,\Phi \right\rangle .  \label{19a}
\end{equation}
In particular 
\begin{equation}
Z(0,f)=1.  \label{19b}
\end{equation}
Then from (\ref{19})-(\ref{19b}) follow that 
\begin{equation}
\left\langle \left| f\right| _{K}^{0},\Phi \right\rangle =\left\langle 1,%
\text{ }\Phi \right\rangle =\frac{Z(0,f)}{Z(-\frac{n}{d},f)}\left\langle
\left| f\right| _{K}^{-\frac{n}{d}},\mathcal{F}\Phi \right\rangle .
\label{19c}
\end{equation}
Therefore (\ref{11a}) is valid for every for $s\in \left\{ \frac{2\pi i}{%
\log q}\mathbb{Z}\right\} $.
\end{proof}

We note that Lemma \ref{lemma1} and \ Theorem \ref{thfe} are valid for
distributions of the form 
\begin{equation*}
\left\langle \chi \left( ac\text{ }f\right) \left| f\right| _{K}^{s},\Phi
\right\rangle =\int\limits_{K^{n}}\Phi \left( x\right) \chi \left( ac\text{ }%
f\left( x\right) \right) \left| f\left( x\right) \right| _{K}^{s}\left|
dx\right| .
\end{equation*}
Functional equations for distributions of type $\chi \left( ac\text{ }%
f\right) \left| f\right| _{K}^{s}$ have been obtain by Igusa \cite{I4}, \
Sato \cite{S1}, Gyoja \cite{G}, and Denef and Meuser \cite{DM}.

\section{Fundamental solutions of \ $p-$adic pseudo-differential equations}

Given a polynomial function $f(x_{1},\ldots ,x_{n})$ with coefficients in $K$
we define a pseudo-differential operator $f(\partial ,\beta )$, $\beta >0$,
that acts on functions in $S(K^{n})$\ by \ 
\begin{eqnarray*}
f(\partial ,\beta )\Phi \left( x\right) &=&\mathcal{F}^{-1}\left( \left|
f\right| _{K}^{\beta }\mathcal{F}\Phi \right) \left( x\right) \\
&=&\int\limits_{K^{n}\setminus f^{-1}\left( 0\right) }\left| f\left(
y\right) \right| _{K}^{\beta }\left( \mathcal{F}\Phi \right) \left( y\right)
\Psi \left( \left[ x,y\right] \right) dy\text{.}
\end{eqnarray*}
Since $\left| f\right| _{K}^{\beta }\mathcal{F}\Phi \mid _{K^{n}\setminus
f^{-1}\left( 0\right) }\in S(K^{n})$, and the Fourier transform is a linear
isomorphism from $S(K^{n})$ onto itself, it holds that $\ f(\partial ,\beta
)\Phi \in S(K^{n})$. Thus $f(\partial ,\beta )$ \ is a linear operator from $%
S(K^{n})$ into $S(K^{n})$. The\ \ operator $f(\partial ,\beta )$ is
continuous and has a self-adjoint extension \ with dense \ domain in $%
L^{2}\left( K^{n}\right) $. We associate to $f(\partial ,\beta )$ the
following \ $p$-adic pseudo-differential equation: 
\begin{equation}
f(\partial ,\beta )u=g\text{, \ }g\in \mathcal{V}\subseteq \mathcal{S}%
(K^{n}).  \label{20}
\end{equation}
A fundamental solution for (\ref{20}) on $\mathcal{V}$ is a distribution $%
E_{\beta }$ such that \ $u=E_{\beta }\ast g$ is a solution. In a recent note 
\cite{Z} the author observed that Atiyah's proof of the existence of a
fundamental solution for a differential operator with constant coefficients
can be adapted to prove the existence of a fundamental solution for (\ref{20}%
) on $\mathcal{S}(K^{n})$.

The following \ Theorem describes explicitly a fundamental solution of (\ref
{20}) as a distribution on $\Delta \left( K^{n}\right) $ when $f(x)$ is a
form of degree $d$ in $n$ variables.

\begin{theorem}
\label{th1}Let $f\left( x\right) $ $\in $ $K\left[ x_{1},\ldots ,x_{n}\right]
\setminus K$ be a form of degree $d$, and $f(\partial ,\beta )$ the $p$-adic
pseudo-differential \ operator with symbol $\left| f\right| _{K}^{\beta }$, $%
\beta >0$. If 
\begin{equation*}
-\beta \notin \cup _{E\in \mathcal{E}}\left\{ -\frac{n_{E}}{N_{E}}\right\}
\cup \left\{ -\frac{n}{d}\right\} \cup \left\{ \gamma \in \mathbb{R}\mid
Z\left( \gamma ,f\right) =0\right\} ,
\end{equation*}
where $\cup _{E\in \mathcal{E}}\left\{ \left( N_{E},n_{E}\right) \right\} $
are the numerical data of $\ $\ a resolution of singularities $(Y,h)$ for $%
f^{-1}(0)$, then the distribution 
\begin{equation*}
E_{\beta }(x)=\left( \frac{1-q^{-d\beta }}{1-q^{-n}}\right) Z\left( -\beta
,f\right) \left\| x\right\| _{K}^{d\beta -n}
\end{equation*}
is a fundamental solution of the $p$-adic pseudo-differential \ equation $%
f(\partial ,\beta )u=g$, \ \ with $g\in \Delta (K^{n}).$
\end{theorem}

\begin{proof}
We set $E_{\beta }=\mathcal{F}^{-1}T$, where $T\in S^{\prime }(K^{n})$
satisfying $\left| f\right| _{K}^{\beta }T=1$. Then $E_{\beta }$ is a
fundamental solution of (\ref{20}). In order to prove the existence of $T$
we proceed as follows. By Theorem \ref{th2} the distribution $\left|
f\right| _{K}^{s}$ \ has a meromorphic continuation to $\mathbb{C}$. Let 
\begin{equation}
\left| f\right| _{K}^{s}=\sum\limits_{m\in \mathbb{Z}}T_{m}\left( s+\beta
\right) ^{m}  \label{for4}
\end{equation}
be the Laurent expansion at $-\beta $ with $T_{m}\in S^{\prime }(K^{n})$ for
all $m$. Since \ the real parts of the poles of $\left| f\right| _{K}^{s}$
are negative rational numbers by Theorem \ref{th2}, $\left| f\right|
_{K}^{s+\beta }=\left| f\right| _{K}^{\beta }\ \left| f\right| _{K}^{s}$\ \
is holomorphic at $s=-\beta $. Therefore $\left| f\right| _{K}^{\beta
}T_{m}=0$ for all $m<0$ and 
\begin{equation}
\left| f\right| _{K}^{s+\beta }=T_{0}\left| f\right| _{K}^{\beta
}+\sum\limits_{m=1}^{\infty }T_{m}\left| f\right| _{K}^{\beta }\left(
s+\beta \right) ^{m}.  \label{for5}
\end{equation}
By using the Lebesgue Lemma and (\ref{for5}) 
\begin{eqnarray}
\lim_{s\rightarrow -\beta }\left\langle \left| f\right| _{K}^{s+\beta },\Phi
\right\rangle &=&\int\limits_{K^{n}}\Phi \left( x\right) \left| dx\right|
=\left\langle 1,\Phi \right\rangle  \notag \\
&=&T_{0}\left| f\right| _{K}^{\beta }.  \label{for6}
\end{eqnarray}
\ Therefore we can take $T=T_{0}$. Now\ since $\left\langle \left| f\right|
_{K}^{s},\Phi \right\rangle $ is a rational function of $q^{-s}$ for every $%
\Phi $ in $S(K^{n})$, and $-\beta $ is not a pole of $\left| f\right|
_{K}^{s}$ it holds that 
\begin{equation*}
\left\langle T_{0},\Phi \right\rangle =\lim_{s\rightarrow -\beta }\text{ }%
\left\langle \left| f\right| _{K}^{s},\Phi \right\rangle ,
\end{equation*}
for every $\Phi $ in $S(K^{n})$, \ i.e., 
\begin{equation}
T_{0}=\lim_{s\rightarrow -\beta }\text{ }\left| f\right| _{K}^{s}.
\label{25}
\end{equation}

In order to find $E_{\beta }$, we compute explicitly $T_{0}$ by using (\ref
{25}), and then $E_{\beta }$ as $\mathcal{F}^{-1}T_{0}$. By Lemma \ \ref
{lemma1} 
\begin{equation}
\left| f\right| _{K}^{s}=\left( \frac{1-q^{-n-ds}}{1-q^{-n}}\right)
Z(s,f)\left\| x\right\| _{K}^{ds},\text{ }s\in \mathbb{C},  \label{28}
\end{equation}
as a distribution on $\Delta \left( K^{n}\right) $. By using (\ref{25}) we
obtain \ from (\ref{28}) that 
\begin{equation}
T_{0}=\left( \frac{1-q^{-n+\beta d}}{1-q^{-n}}\right) Z(-\beta ,f)\left\|
x\right\| _{K}^{-\beta d}  \label{29}
\end{equation}
as a distribution on $\Delta \left( K^{n}\right) $. If $\beta \neq \frac{n}{d%
}$, using Proposition (\ref{prop1}) and the fact that the Fourier transform
is an isomorphism on $\Delta \left( K^{n}\right) $, we obtain that $\mathcal{%
F}^{-1}T_{0}$ is equal to \ 
\begin{equation*}
E_{\beta }\left( x\right) =\left( \frac{1-q^{-d\beta }}{1-q^{-n}}\right)
Z(-\beta ,f)\left\| x\right\| _{K}^{\beta d-n},\text{ if }\beta \neq \frac{n%
}{d},\text{ }
\end{equation*}
as a distribution on $\Delta \left( K^{n}\right) $. Finally, the condition $%
Z(-\beta ,f)\neq 0$ implies $E_{\beta }\neq 0$.
\end{proof}

\subsection{Remarks\label{reamark}}

Let $E$ be in $\mathcal{S}^{\prime }(K^{n})$, and $g$ a complex valued
function defined on $X\subseteq K^{n}$ having 
\begin{equation*}
g(x)=\sum\limits_{i=0}^{\infty }a_{k}\varphi _{k}\left( x\right) \text{, \ }%
x\rightarrow x_{0},
\end{equation*}
as an asymptotic expansion as $x$ tends to $x_{0},$ here $x_{0}$\ is a\
limit point of $X$. If 
\begin{equation*}
\left\langle E,\Phi \right\rangle =\int\limits_{K^{n}}g\left( x\right) \Phi
\left( x\right) \left| dx\right| ,
\end{equation*}
for any $\Phi \in \mathcal{S}(K^{n})$ whose support is contained in a
sufficiently small ball around $x_{0}$, then we shall say \ that $E$ has \
an asymptotic expansion \ as $x$ tends to $x_{0}$, and write 
\begin{equation*}
E\left( x\right) =\sum\limits_{i=0}^{k}a_{k}\varphi _{k}\left( x\right) +%
{\LARGE O}\left( \varphi _{k+1}\left( x\right) \right) \text{, \ }%
x\rightarrow x_{0}.
\end{equation*}

\begin{corollary}
\label{corola1}With the hypothesis of Theorem \ref{th1}, it holds that (I) $%
E_{\beta }(x)=O\left( \left\| x\right\| _{K}^{d\beta -n}\right) $ as $%
\left\| x\right\| _{K}$ $\rightarrow 0$; (II) $E_{\beta }(x)=O\left( \left\|
x\right\| _{K}^{-d\beta }\right) $ as $\left\| x\right\| _{K}$ $\rightarrow
\infty $. Moreover, the fundamental solution $E_{\beta }(x)$ is non-singular
at the origin if $\beta >\frac{n}{d}$.
\end{corollary}

\begin{proof}
The first part follows directly from Theorem \ref{th1}, and Remark \ref
{reamark}\ (3). The second part follows from the fact that 
\begin{equation*}
\left\langle E_{\beta }(x),\Omega _{-\ell }\right\rangle =\left\langle
E_{\beta }(x),\mathcal{F}\Omega _{\ell }\right\rangle =\left\langle \left( 
\frac{1-q^{d\beta -n}}{1-q^{-n}}\right) Z\left( -\beta ,f\right) \left\|
x\right\| _{K}^{-d\beta },\Omega _{\ell }\right\rangle .
\end{equation*}
\end{proof}

\subsection{Elliptic pseudo-differential operators}

A quadratic form 
\begin{equation*}
h(x_{1},\ldots ,x_{n})\in \mathbb{Q}_{p}\left[ x_{1},\ldots ,x_{n}\right] 
\text{, }p\neq 2,
\end{equation*}
is called \ \textit{elliptic} (or anisotropic) if it satisfies 
\begin{equation}
h(x_{1},\ldots ,x_{n})\neq 0\text{ if \ }\left| x_{1}\right| _{\mathbb{Q}%
_{p}}+\ldots +\left| x_{n}\right| _{\mathbb{Q}_{p}}\neq 0.  \label{35}
\end{equation}
Other quadratic forms are called \textit{isotropic}. A pseudo-differential
operator $h(\partial ,\beta )$ with symbol $\left| h\right| _{K}^{\beta }$
satisfying (\ref{35}) is called an elliptic operator. It is well-known that
there no exist anisotropic quadratic forms if $n\geqslant 5$. The following
table \ shows all the anisotropic quadratic forms up to linear isomorphism.

\begin{equation}
\begin{tabular}{|l|l|}
\hline
$n$ & \ \ \ \ \ \ \ \ \ \ \ \ \ \ \ \ \ \ \ \ Quadratic Forms \\ \hline
$2$ & $h(x_{1},x_{2})=x_{1}^{2}-\tau x_{2}^{2},\tau \in \mathbb{Q}%
_{p}\setminus \mathbb{Q}_{p}\mathbb{Q}_{p},$ $\tau =\epsilon ,\tau =p,\tau
=\epsilon p,\epsilon \in \mathbb{Z}_{p}^{\times }$ \\ \hline
$3$ & 
\begin{tabular}{r}
$
\begin{array}{c}
h(x_{1},x_{2},x_{3})=p\epsilon _{1}x_{1}^{2}+\epsilon _{2}x_{2}^{2}+\epsilon
_{3}x_{3}^{2}\text{, }\epsilon _{1},\epsilon _{2},\epsilon _{3}\in \mathbb{Z}%
_{p}^{\times }\text{,} \\ 
\text{ \ }\epsilon _{2}x_{2}^{2}+\epsilon _{3}x_{3}^{2}\neq 0\text{ if \ }%
\left| x_{1}\right| _{\mathbb{Q}_{p}}+\left| x_{2}\right| _{\mathbb{Q}%
_{p}}\neq 0
\end{array}
$%
\end{tabular}
\\ \hline
$4$ & $%
h(x_{1},x_{2},x_{3},x_{4})=x_{1}^{2}-sx_{2}^{2}-px_{3}^{2}+spx_{4}^{2} $, $%
s\in \mathbb{Z}$, with $\left( \frac{s}{p}\right) =-1$ \\ \hline
\end{tabular}
\label{table}
\end{equation}

Our next goal is \ to determine a fundamental solution for an \ elliptic
operator on $\Delta (K^{n})$. The first step is to calculate $Z(s,h)$. This
calculation can be easily accomplished by using the $p-$adic stationary
phase formula. This method introduced by Igusa \cite{I3} permits \ the
calculation of the local zeta function for a wide \ class of polynomials 
\cite[and the references therein]{I3}, \cite{Z2}, \cite{Z3}, \cite{Z4}.

If $h\left( x\right) \in R_{K}\left[ x_{1},\ldots ,x_{n}\right] \setminus $ $%
\frak{p}R_{K}$ $\left[ x_{1},\ldots ,x_{n}\right] $, we denote by $\overline{%
h}\left( x\right) $ its reduction modulo $\frak{p}R_{K}$.

\begin{proposition}[{\protect\cite[Proposition 10.2.1]{I1}}]
\label{prop2}If 
\begin{equation*}
h\left( x\right) \in R_{K}\left[ x_{1},\ldots ,x_{n}\right] \setminus \frak{p%
}R_{K}\left[ x_{1},\ldots ,x_{n}\right]
\end{equation*}
is a homogeneous polynomial of degree $d$ such that $\overline{h}\left( 
\overline{a}\right) =\frac{\partial \overline{h}}{\partial x_{i}}\left( 
\overline{a}\right) =0$ for $1\leq i\leq n$ implies $\overline{a}=0$, then 
\begin{equation*}
Z(s,h)=\frac{\left( 1-q^{-n}N\right) +\left( q^{-n-1}+q^{-n}\left(
N-1\right) -q^{-1}\right) q^{-s}}{\left( 1-q^{-1-s}\right) \left(
1-q^{-n-ds}\right) },
\end{equation*}
where $N$ denotes the number of zeros of \ $\overline{f}\left( x\right) $ in 
$\mathbb{F}_{q}^{n}$.
\end{proposition}

We shall identify an elliptic quadratic form with one of the polynomials
listed in \ table (\ref{table}). Then as a consequence of the previous
Proposition we obtain the following result.

\begin{corollary}
\label{cor1}If $h(x)\in \mathbb{Z}_{p}\left[ x\right] $ is an elliptic
quadratic form in $n$ variables, then \ $Z(s,h)=\frac{1-q^{-n}}{1-q^{-n-2s}}%
. $
\end{corollary}

The following result describes explicitly a fundamental solution of $%
h(\partial ,\beta )u=g$, $g\in \Delta \left( K^{n}\right) $; this result
follows from Theorem \ref{th1} and Corollary \ref{cor1}.

\begin{theorem}
\label{th3}Let $h\left( x\right) $ $\in $ $\mathbb{Z}_{p}\left[ x_{1},\ldots
,x_{n}\right] $ be an elliptic quadratic form, and $h(\partial ,\beta )$ the 
$p$-adic pseudo-differential \ operator with symbol $\left| h\right|
_{K}^{\beta }$, $\beta >0$. If $\beta \neq \frac{n}{2}$, then the
distribution 
\begin{equation*}
E_{\beta }(x)=\left( \frac{1-p^{-2\beta }}{1-p^{-n+2\beta }}\right) \left\|
x\right\| _{\mathbb{Q}_{p}}^{2\beta -n}
\end{equation*}
is a fundamental solution of the $p$-adic pseudo-differential \ equation $%
h(\partial ,\beta )u=g$, \ \ with $g\in \Delta (K^{n}).$
\end{theorem}

Theorem \ref{th3} is valid for any \ finite extension $K$ of $\mathbb{Q}_{p}$%
. In \cite{Koch2}, \cite[Chap. 2]{Koch1}\ Kochubei calculated explicitly the
\ fundamental solutions for the elliptic operators. The restriction of these
distributions to the space $\Delta (K^{n})$ coincide \ with the
distributions given in Theorem \ref{th3} when $\beta \neq \frac{n}{2}$.

\section{\label{sect}$p-$adic Green Functions}

A fundamental solution $G_{\lambda }$ of the pseudo-differential 
\begin{equation*}
\left( f\left( \partial ,\beta \right) +\lambda \right) u=g,\text{ \ \ \ }%
\beta ,\lambda \in \mathbb{R}\text{, }\beta ,\lambda >0,\text{ \ }g\in 
\mathcal{S}(K^{n})
\end{equation*}
is called a $p-$\textit{adic Green function}. Since $\left| f\left( x\right)
\right| _{K}^{\beta }+\lambda \neq 0$, for every $x$ in $K^{n}$, \ the
distribution \ 
\begin{equation*}
G_{\lambda }=\mathcal{F}^{-1}\left( \frac{1}{\left| f\right| _{K}^{\beta
}+\lambda }\right)
\end{equation*}
is a Green function. We shall say that \ $G_{\lambda }$ is the \textit{Green
function associated} to the operator $f\left( \partial ,\beta \right) $.

\subsection{Integration on the fibers}

Let \ $g(x)$ be a non-constant polynomial with coefficients in $K$. Let $%
C_{g}$ and $S_{g}=g(C_{g})$ denote the set of critical points and the set of
critical values of the mapping $g:K^{n}\rightarrow K$. In the case in which $%
g$ a homogeneous polynomial $S_{g}=\left\{ 0\right\} $ (see e.g. 
\cite[Theoren 2.5.1]{I1}). For any $z\in K^{n}\setminus \left\{ 0\right\} $
we define $\left| \frac{dx}{dg}\right| $ to be the residue of the measure $%
\left| dx\right| $ along the fiber $g^{-1}\left( z\right) $, and 
\begin{equation}
F_{\Phi }\left( z\right) =\int\limits_{g^{-1}\left( z\right) }\Phi \left(
x\right) \left| \frac{dx}{dg}\right| ,
\end{equation}
where \ $\Phi \in \mathcal{S}(K^{n})$ is a fixed function. If $\Phi $\ is
the characteristic function of $R_{K}^{n}$ we use $F\left( z\right) $
instead of $F_{\Phi }\left( z\right) $. The function $F_{\Phi }\left(
z\right) $ is locally constant \ on $K\setminus \left\{ 0\right\} $, and \
satisfies 
\begin{equation}
\int\limits_{K^{n}}\Phi \left( z\right) \left| dz\right|
=\int\limits_{K\setminus \left\{ 0\right\} }F_{\Phi }\left( z\right) \left|
dz\right|  \label{48}
\end{equation}
for every $\Phi $ in $\mathcal{S}(K^{n})$ (cf. \cite[Lemma 8.3.2]{I1}). The
following Lemma, that follows from (\ref{48}), will be used later on.

\begin{lemma}[{cf. \protect\cite[Theorem 8.4.1]{I1}}]
\label{th5}Let $g(x)\in K\left[ x_{1},\ldots ,x_{n}\right] \setminus K$ be a
non-\-con\-stant polynomial such that $C_{g}$ is contained in $g^{-1}(0)$.
Then 
\begin{equation}
Z_{\Phi }(s,g)=\int\limits_{K\setminus \left\{ 0\right\} }F_{\Phi }\left(
z\right) \left| z\right| _{K}^{s}\left| dz\right| ,\text{ \ }\func{Re}(s)>0.
\label{49}
\end{equation}
\end{lemma}

\subsection{ Asymptotics of $G_{\protect\lambda }(x)$ as $\left\| x\right\|
_{K}\rightarrow \infty $}

\begin{theorem}
\label{th4}Let $f\left( x\right) $ $\in $ $R_{K}\left[ x_{1},\ldots ,x_{n}%
\right] \setminus K$ be a form of degree $d$, and $f(\partial ,\beta )$ the $%
p$-adic pseudo-differential \ operator with symbol $\left| f\right|
_{K}^{\beta }$, $\beta >0.$ The Green function $G_{\lambda }$ corresponding
to $f(\partial ,\beta )$ admits the asymptotic expansion 
\begin{equation*}
G_{\lambda }(x)=\left( \frac{1}{1-q^{-n}}\right) \sum\limits_{m=1}^{\infty }%
\frac{\left( -1\right) ^{m-1}\left( 1-q^{d\beta m}\right) }{\lambda ^{m+1}}%
Z\left( \beta m,f\right) \left\| x\right\| _{K}^{-d\beta m-n}
\end{equation*}
as \ $\left\| x\right\| _{K}\rightarrow \infty $. In particular $G_{\lambda
}(x)$\ satisfies 
\begin{equation*}
G_{\lambda }(x)=\left( \frac{1}{1-q^{-n}}\right) \frac{\left( 1-q^{d\beta
}\right) }{\lambda ^{2}}Z\left( \beta ,f\right) \left\| x\right\|
_{K}^{-d\beta -n}+{\LARGE O}\left( \frac{1}{\lambda ^{3}}\left\| x\right\|
_{K}^{-2d\beta -n}\right)
\end{equation*}
as \ $\left\| x\right\| _{K}\rightarrow \infty $.
\end{theorem}

\begin{proof}
The asymptotic behavior of $G_{\lambda }$ at infinity can be studied by
considering the action of $G_{\lambda }$ on functions \ of type $\Omega
_{-l} $, $\ $the characteristic function of the ball $\left(
P_{K}^{-l}\right) ^{n} $, as $l\rightarrow \infty $ (see Remark \ref{reamark}%
). By definition

\begin{eqnarray}
\left\langle G_{\lambda },\Omega _{-l}\right\rangle &=&\left\langle \frac{1}{%
\left| f\right| _{K}^{\beta }+\lambda },\mathcal{F}^{-1}\Omega
_{-l}\right\rangle =\left\langle \frac{1}{\left| f\right| _{K}^{\beta
}+\lambda },q^{nl}\Omega _{l}\right\rangle  \notag \\
&=&\int\limits_{R_{K}^{n}}\frac{\left| dz\right| }{q^{-ld\beta }\left|
f\left( z\right) \right| _{K}^{\beta }+\lambda },  \label{62}
\end{eqnarray}
and by using integration on the fibers, (\ref{62}) can be rewritten as 
\begin{equation}
\left\langle G_{\lambda },\Omega _{-l}\right\rangle
=\int\limits_{R_{K}\setminus \left\{ 0\right\} }\frac{F(t)}{q^{-ld\beta
}\left| t\right| _{K}^{\beta }+\lambda }\left| dt\right| \text{.}  \label{64}
\end{equation}
By applying the asymptotic expansion 
\begin{equation*}
\frac{1}{1+y}=\sum\limits_{m=1}^{\infty }\left( -1\right) ^{m-1}y^{n},\text{
as }y\rightarrow 0,
\end{equation*}
in \ (\ref{64}) we obtain the following asymptotic expansion for $%
\left\langle G_{\lambda },\Omega _{-l}\right\rangle $: 
\begin{equation}
\left\langle G_{\lambda },\Omega _{-l}\right\rangle
=\sum\limits_{m=1}^{\infty }\left( -1\right) ^{m-1}\frac{q^{-ld\beta m}}{%
\lambda ^{m+1}}\int\limits_{R_{K}\setminus \left\{ 0\right\} }\left|
t\right| _{K}^{\beta m}F(t)\left| dt\right| ,\text{ }l\rightarrow \infty .
\label{65}
\end{equation}

By using Lemma \ \ref{th5} and (\ref{13}) in (\ref{65}), we obtain that 
\begin{eqnarray*}
\left\langle G_{\lambda },\Omega _{-l}\right\rangle
&=&\sum\limits_{m=1}^{\infty }\left( -1\right) ^{m-1}\frac{q^{-ld\beta m}}{%
\lambda ^{m+1}}Z\left( \beta m,f\right) = \\
&&\sum\limits_{m=1}^{\infty }\left( -1\right) ^{m-1}\frac{Z\left( \beta
m,f\right) }{\lambda ^{m+1}}\left\langle \frac{1-q^{-n-d\beta m}}{1-q^{-n}}%
\left\| x\right\| _{K}^{d\beta m},q^{nl}\Omega _{-l}\right\rangle ,
\end{eqnarray*}
as $l\rightarrow \infty $, i.e., 
\begin{equation}
\left\langle G_{\lambda },\Omega _{-l}\right\rangle
=\sum\limits_{m=1}^{\infty }\left( -1\right) ^{m-1}\frac{Z\left( \beta
m,f\right) }{\lambda ^{m+1}}\left\langle \mathcal{F}^{-1}\left( \frac{%
1-q^{-n-d\beta m}}{1-q^{-n}}\left\| x\right\| _{K}^{d\beta m}\right) ,\Omega
_{-l}\right\rangle .  \label{67}
\end{equation}
From (\ref{67}) by using Proposition \ref{prop1} we obtain that 
\begin{equation}
G_{\lambda }\left( x\right) =\sum\limits_{m=1}^{\infty }\left( -1\right)
^{m-1}\frac{\left( 1-q^{d\beta m}\right) Z\left( \beta m,f\right) }{\left(
1-q^{-n}\right) \lambda ^{m+1}}\left\| x\right\| _{K}^{-d\beta m-n},\text{ }%
\left\| x\right\| _{K}\rightarrow \infty .\text{ }  \label{68}
\end{equation}
The second part \ follows directly from (\ref{68}).
\end{proof}

\subsection{Remarks}

\begin{enumerate}
\item  The previous Theorem is valid for a twisted operator $f(\partial
,\beta ,\chi )$, in this case it is necessary to change $Z\left( \beta
m,f\right) $ by $Z(\beta m,\chi ,f)$ in \ \ the statement of Theorem \ref
{th4}.

\item  Kochubei \ has studied the asymptotics of the Green functions
associated with elliptic operators (see e.g. \cite[sect. 2.8]{Koch1}) at
infinity and the origin. The asymptotics obtained by Kochubei at infinity
can be recovered from \ Theorem \ref{th4}.

\item  Sato's \ asymptotic expansion for $G_{\lambda }$, \cite{S}, is only
valid for \ forms that are invariants of prehomogeneous vector spaces.
\end{enumerate}


\begin{thebibliography}{99}
\bibitem{A}  Atiyah, M. F., Resolution of singularities and division of
distributions, Comm. Pure Appl. Math. 23 (1970), 145-150.

\bibitem{DM}  Denef J., and Meuser D., \ A functional equation of Igusa's
local zeta function, Amer. J. Math., 109 (1987), 991-1008.

\bibitem{H}  Hironaka H., Resolution of singularities of an algebraic
variety over a field of characteristic zero, Ann. Math., 79 (1964), 109-326.

\bibitem{G}  Gyoja A., Lectures on the functional equation satisfied by the $%
p-$adic local zeta functions of reductive prehomogeneous vector spaces,
(unpublished) lectures at the Johns Hopkins University, 1993.

\bibitem{I1}  Igusa Jun-Ichi, An introduction to the theory of local zeta
functions, AMS /IP studies in \ advanced mathematics, v. 14, 2000.

\bibitem{I2}  Igusa J.-I., \ Complex powers and asymptotic expansions, I
Crelles J. Math., 268/269 (1974), 110-130; II, ibid., 278/279 (1975),
357-368.

\bibitem{I3}  Igusa, J.-I., A stationary phase formula for $p-$adic
integrals \ and its applications, in Algebraic Geometry and its
Applications, Springer-Verlag (1994), 175-194.

\bibitem{I4}  Igusa, J. -I., Some results on $p-$adic complex powers, Amer.
J. Math. 106 (1984), 1013-1032.

\bibitem{J}  Jang Y, An asymptotic expansion of the $p-$adic Green function,
Tohoku Math. J., 50 (1998), 229-242.

\bibitem{Koch1}  Kochubei A. N., Pseudodifferential equations and
stochastics over non-archimedean fields, Marcel Dekker, 2001.

\bibitem{Koch2}  Kochubei A. N., Fundamental solutions of
pseudo-differential equations associated with $p-$adic quadratic forms,
Izvestiya Math., 62 (1998), 1169-1188.

\bibitem{K2}  Kochubei A. N., On $p-$adic Green functions, Theor. Math.
Phys., 96 (1993), 854-865.

\bibitem{K3}  Kochubei A. N., on the asymptotics of $p-$adic Green
functions, Proc. Steklov Inst. Math. 203 (1994/95), 105-113.

\bibitem{Khr}  Khrennikov A. , Fundamental solutions over the field \ of $p-$%
adic numbers, St. Peterburgh Math. J., 4 (1993), 613-628.

\bibitem{S}  Sato Fumihiro , $p-$adic Green functions and zeta functions,
Commentarii Mathematici Universitatis Sancti Pauli, 51 (2002), 79-97.

\bibitem{S1}  Sato F., \ On functional equations of zeta distributions, Adv.
\ Studies in Pure Math., 15 (1989), 465-508.

\bibitem{T}  Taibleson M. H., \ Fourier analysis on local fields,
Mathematical Notes, Vol. 15, \ Princeton University Press, 1975.

\bibitem{V}  Vladimirov V. S., On the spectrum of some pseudo-differential
operators over $p-$adic number field, Algebra and Analysis, 2 (1990),
107-124.

\bibitem{VVZ}  Vladimirov V. S., Volovich I. V., and Zelenov E. I., $p-$adic
Analysis \ and mathematical physics, World Scientific, Singapore, 1994.

\bibitem{W}  Weil A. , Sur la formule de Siegel dans le th\'{e}orie des
groupes classiques, \ Acta Math. 113 (1965), 1-87.

\bibitem{Z}  Zuniga-Galindo W.\ A., Fundamental solutions of
pseudo-differential operators over $p-$adic fields, Rend. Sem. Mat. Univ.
Padova. 109 (2003), 241--245.

\bibitem{Z2}  Zuniga-Galindo W. A., Igusa's local zeta functions of
semiquasihomogeneous polynomials, Trans. Amer. Math. Soc. 353, (2001),
3193-3207.

\bibitem{Z3}  Zuniga-Galindo W. A., Local zeta functions and Newton
Polyhedra, Nagoya Math. J., Vol 172 (2003), 31-58.

\bibitem{Z4}  Zuniga-Galindo W.A., Local zeta function for non-degenerate
homogeneous mappings, to appear in Pac. J. Math.
\end{thebibliography}
\end{document}